\documentclass[prl,twocolumn,amsmath,amssymb,superscriptaddress,citeautoscript,showpacs,longbibliography]{revtex4-1}

\usepackage{graphicx,color}
\usepackage{braket}
\usepackage[colorlinks,urlcolor=blue,citecolor=blue]{hyperref}

\newcommand{\Rcal}{\mathcal{R}}
\newcommand{\Qcal}{\mathcal{Q}}
\newcommand{\Tcal}{\mathcal{T}}

\begin{document}
\title{Matrix product ansatz for Fermi fields in one dimension}

\author{Sangwoo S.~\surname{Chung}}
\affiliation{Department of Physics, University of Cincinnati, Cincinnati, Ohio
45221-0011, USA}
\author{Kuei \surname{Sun}}
\affiliation{Department of Physics, The University of Texas at Dallas, Richardson,
Texas 75080-3021, USA}
\affiliation{Department of Physics, University of Cincinnati, Cincinnati, Ohio
45221-0011, USA}
\author{C.~J.~\surname{Bolech}}
\affiliation{Department of Physics, University of Cincinnati, Cincinnati, Ohio
45221-0011, USA}

\begin{abstract}
We present an implementation of a continuous matrix product state for two-component fermions in one dimension.  We propose a construction of variational matrices with an efficient parametrization that respects the translational symmetry of the problem (without being overly constraining) and readily meets the regularity conditions that arise from removing the ultraviolet divergences in the kinetic energy.  We test the validity of our approach on an
interacting spin-1/2 system and
observe that the ansatz correctly predicts the ground-state magnetic properties
for the attractive spin-1/2 Fermi gas, including the phase-oscillating pair
correlation function in the partially polarized regime.
\end{abstract}

\maketitle

An exact description of an arbitrary many-body quantum state can be a formidable task, as the number of distinct states grows exponentially on the number of particles or system volume.  Nevertheless, accurate approximations of low-energy states are often feasible, as the vast majority of the quantum states in the entire Hilbert space need not be considered \cite{Eisert2010}.  Among numerical treatments of one-dimensional (1D) strongly correlated problems, the density matrix renormalization group (DMRG) algorithm \cite{White1992,*White1993,*Schollwock2005} for lattice systems has been particularly notable for its optimal efficiency.  Moreover, the DMRG procedure is essentially equivalent to an optimization of a matrix product state (MPS) \cite{Affleck1987,*Ostlund1995,*Cirac2009,*Schollwock2011} --- a natural and effective form of parametrization of variational states for systems with limited entanglement due to short-range interactions.

Recently, a coherent-state extension of the MPS to quantum fields in a 1D continuum (called cMPS) \cite{Verstraete2010, Maruyama2010, Osborne2010, Haegeman2010, Rispler2012, Haegeman2013a, Draxler2013,  Hubener2013, Quijandria2014, Stojevic2015, Haegeman2015} has been pioneered, and it was successful in correctly predicting the ground-state energies \cite{Verstraete2010} and dispersion relations \cite{Draxler2013} of interacting bosons.  For fermionic systems, on the other hand, the cMPS has been tested on a 1D relativistic model related to quantum chromodynamics \cite{Haegeman2010}, but its success for nonrelativistic fermions --- which have a wide range of applications in condensed matter, quantum information, as well as cold-atom systems --- has not been demonstrated thus far.  In particular, as recent experimental techniques \cite{Bloch2008,*Giorgini2008} have enabled the physical realization of ultracold 1D Fermi gas systems \cite{Liao2010,Moritz2005,Pagano2014,Guan2013} with a tunable interaction, particle density, and spin polarization, a cMPS ansatz that can reliably investigate those systems would be of significant interest.

This Rapid Communication extends the previous developments and proposes an
implementation of a cMPS ansatz for two-component fermions.  Although some important insights on the structure of a fermionic ansatz have been noted previously \cite{Haegeman2010,Rispler2012,Haegeman2013a}, a subsequent work that fully develops a more efficient and explicit scheme that i) readily satisfies the regularity conditions for a finite kinetic energy, as well as ii) being able to capture spontaneous translational symmetry breaking properties, is much needed to advance the subject further.  We shall show that our cMPS ansatz is able to faithfully portray the ground-state magnetic properties of a
spin-1/2 Fermi gas described by the Gaudin-Yang Hamiltonian \cite{Gaudin1967,*Yang1967}, in agreement with the exact solutions obtained from the thermodynamic Bethe ansatz \cite{Takahashi1971,Takahashi1999}, including the distinct phases for the attractive system and spontaneous translational symmetry breaking properties such as inhomogeneous (oscillatory-phase) superconductivity in the partially polarized regime.

\begin{figure}[b]
\includegraphics[width=\columnwidth]{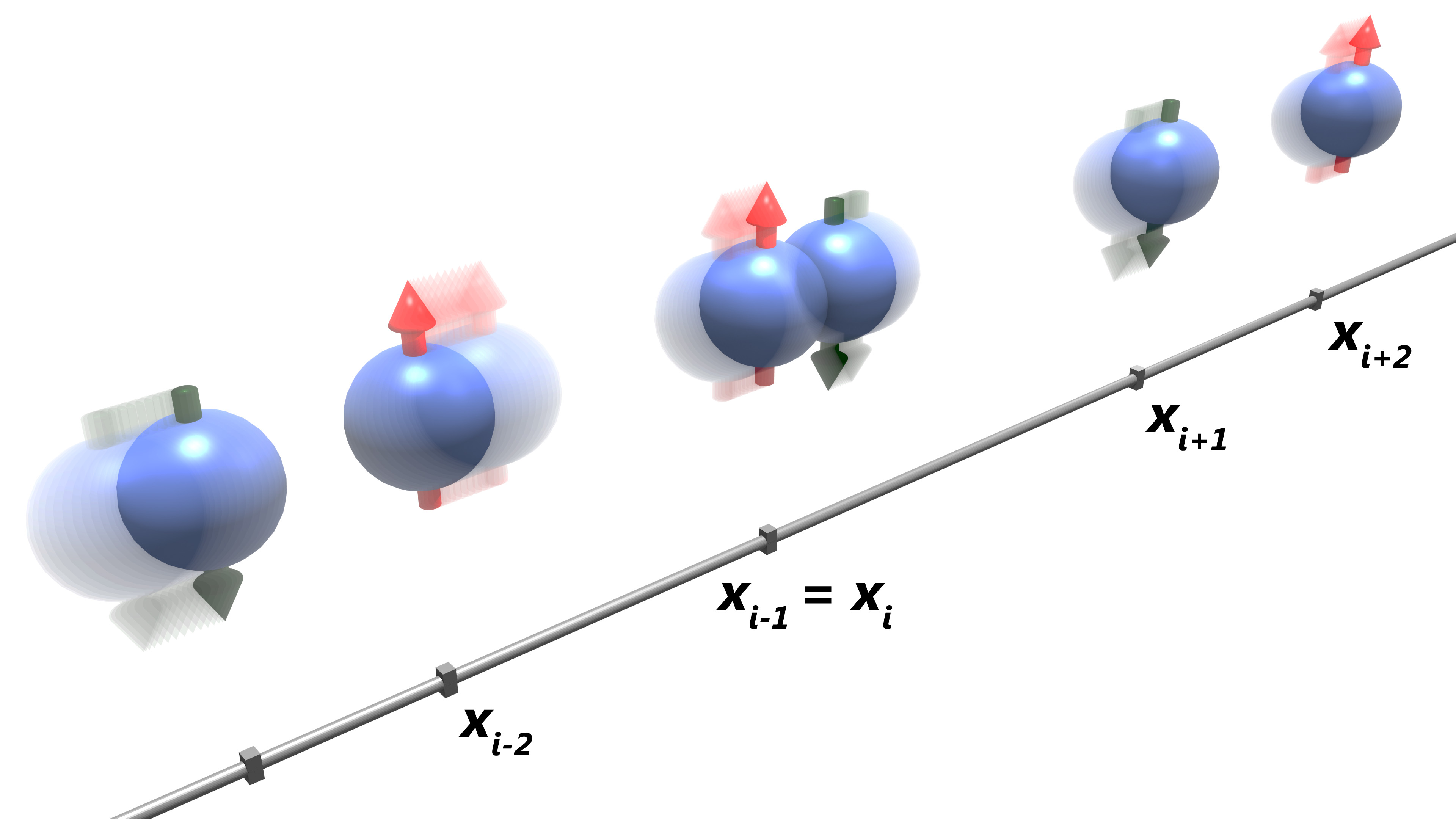}
\caption{\label{fig:cmps}
(Color online) Artistic rendition of the cMPS in Eq.~(\ref{eq:cmps}), that when explicitly expanded, can alternatively be understood as a superposition of position eigenstates, each state having some fixed number of particles that ranges from 1 to $\infty$.  The probability amplitude for the particular position eigenstate illustrated above for $x_{i-1}=x_i$ is given by $\phi_{\cdots\uparrow\uparrow\downarrow\downarrow\cdots}(...,x_{i-2},x_{i-1},x_{i},x_{i+1},...)\allowbreak=\mathrm{Tr}[\cdots\Rcal_{\uparrow}(x_{i-2})\allowbreak u(x_{i-2},x_{i-1})\allowbreak \Rcal_{\uparrow}(x_{i-1})\allowbreak \Rcal_{\downarrow}(x_{i})\allowbreak u(x_{i},x_{i+1})\allowbreak \Rcal_{\downarrow}(x_{i+1})\cdots]$, where $u(x,y)\allowbreak=\mathcal{P}\mathrm{exp}\left[\int_{x}^{y}dx\Qcal(x)\right]$ can be viewed as a free propagator and $\Rcal_{\sigma}(x)$ as a vertex that creates a particle of type $\sigma$ at position $x$ \cite{Haegeman2013a}. }
\end{figure}

A cMPS for spin-1/2 fermions (see Fig.~\ref{fig:cmps}), for a system
length of $L$ and periodic boundary conditions, has the general form \cite{Haegeman2010}
\begin{eqnarray}
\left|\chi\right\rangle & = &
\mathrm{Tr_{aux}}[\mathcal{P}e^{\int_{0}^{L}dx\left[\Qcal(x)\otimes \hat{I}+\sum_{\sigma}\Rcal_{\sigma}(x)\otimes\hat{\psi}_{\sigma}^{\dagger}(x)\right]}]\left|\Omega\right\rangle,\label{eq:cmps}
\end{eqnarray}
where $\Qcal(x)$, $\Rcal{_\sigma}(x)\in\mathbb{C}^{D\times D}$ and act on a
$D$-dimensional auxiliary space, $D$ is called the \emph{bond dimension}, $\sigma$ is the index for spins $\uparrow$ and
$\downarrow$, $\hat{I}$ is the identity operator on the Fock space, $\mathrm{Tr_{aux}}$ is a trace over the auxiliary space,
$\mathcal{P}$exp is a path-ordered exponential, and $\left|\Omega\right\rangle$ is
the Fock-space vacuum state. The Fock-space creation and annihilation field operators obey the
anticommutation relation $\{
\hat{\psi_{\sigma}}(x),\hat{\psi}^{\dagger}_{\sigma'}(x')\}
=\delta_{\sigma\sigma'}\delta(x-x')$. 

In order to illustrate our implementation of the cMPS ansatz, we shall consider the Gaudin-Yang Hamiltonian \cite{Gaudin1967,*Yang1967}, where spin-1/2 fermions interact via a contact potential and can be written as
\begin{eqnarray}
H_{\mathrm{GY}} & = &
\int_{0}^{L}dx\sum_{\sigma=\uparrow,\downarrow}\left(\partial_{x}\hat{\psi}_{\sigma}^{\dagger}\partial_{x}\hat{\psi}_{\sigma}-2\Delta\hat{\psi}_{\sigma}^{\dagger}\hat{\psi}_{\bar{\sigma}}^{\dagger}\hat{\psi}_{\bar{\sigma}}\hat{\psi}_{\sigma}\right),\label{eq:GY}\end{eqnarray}
where $\bar{\sigma}$ denotes the conjugate spin to $\sigma$, and the interaction strength is $\Delta<0$ and $\Delta>0$ for repulsive and
attractive interactions, respectively.  The exact wave function for the two-particle spin-singlet sector of this translationally invariant system
with periodic boundary conditions can be obtained using the coordinate
Bethe ansatz as \cite{Takahashi1999}
\begin{eqnarray}
\phi_s\left(x_{1},x_{2}\right) & = & A_s e^{\frac{i}{2}\left(k_{1}+k_{2}\right)\left(x_{1}+x_{2}\right)}\nonumber \\
 &  & \times\left(e^{\frac{i}{2}\left(k_{2}-k_{1}\right)\left|x\right|}+S_se^{-\frac{i}{2}\left(k_{2}-k_{1}\right)\left|x\right|}\right),\label{eq:bethe}
\end{eqnarray}
where $k_{j}$ is the quasimomentum of the $j$th particle (obeying the corresponding Bethe ansatz equation), $x\equiv x_{2}-x_{1}$, $S_s=\left(k_{1}-k_{2}-2i\Delta\right)^{-1}\left(k_{1}-k_{2}+2i\Delta\right)$,
and $A_s$ is an overall amplitude. By constructing a cMPS with
\begin{eqnarray}
\Qcal(x) &=& \frac{i}{2}\mathrm{diag}\left(
q_{-}-q_{+}\quad k_2-k_1\quad k_1-k_2\quad q_{+} - q_{-}\right),\nonumber\\ 
\Rcal_{\uparrow}(x) &=&\frac{e^{iq_{+}x}}{2}\left(\begin{array}{cc}
\sigma_{+} & 2I_{2}\\
0 & -\sigma_{+}
\end{array}\right),\label{eq:twosector}\\
\Rcal_{\downarrow}(x) &=&\frac{A_s(1+S)e^{iq_{-}x}}{4i\sin[(k_{2}-k_{1})L/2]}\left(\begin{array}{cc}
\sigma_{-} & 0\\
-2I_{2} & -\sigma_{-}
\end{array}\right),\nonumber
\end{eqnarray}
where $I_d$ is a $d\times d$ identity matrix, $\sigma_{\pm}=\sigma_{x}\pm i\sigma_{y}$ with Pauli matrices $\sigma_{x,y,z}$, and $q_{\pm}=\left(k_{1}+k_{2}\right)/2\pm n\pi/L$ with $n\in\mathcal{\mathfrak{\mathbb{Z}}}$ (its parity a function of $k_2-k_1$), we
recover the exact wave function in Eq.~(\ref{eq:bethe}) after projecting the cMPS
onto the two-particle sector.

We remark the phase modulation present on the $\mathcal{R}_{\sigma}$ matrices in Eq.~(\ref{eq:twosector}), while the Hamiltonian is translationally invariant, is reminiscent of the general result for Bloch (or Floquet) states.  In fact, it can be shown that $e^{iq_{\sigma}x}$ is the most general form of modulation on otherwise (spatially) constant $\mathcal{R}_{\sigma}$ matrices that preserves the continuous translational symmetry of the system. Based on this insight, we depart from the extended practice of taking both $\Qcal(x)$ and $\Rcal_{\sigma}(x)$ to be independent of coordinates, the so-called uniform ansatz or uMPS \cite{Milsted2013,*Haegeman2013b}, and propose a phase modulation on $\Rcal_{\sigma}(x)$.  Henceforth, $\Qcal(x)=Q$ and $\Rcal_{\sigma}(x)=R_{\sigma}e^{iq_{\sigma}x}$, where $Q$ and $R_{\sigma}$ are $D\times D$ matrices that are independent of position, and $q_{\sigma}$ are real variational parameters, unconstrained in the thermodynamic limit.  We note that despite
this modulation of $\mathcal{R}_{\sigma}$,  
$\Tcal(x)\equiv \Qcal(x)\otimes I+I\otimes\bar{\Qcal}(x)+\Rcal_{\uparrow}(x)\otimes\bar{\Rcal}_{\uparrow}(x)+\Rcal_{\downarrow}(x)\otimes\bar{\Rcal}_{\downarrow}(x)\equiv T$
remains translationally invariant (the bars denote complex conjugation of matrix entries).  Some of the correlators and observables that we will use below are
\begin{eqnarray}
\bigl\langle\hat{\psi}_{\sigma'}^{\dagger}(x+\delta x)\hat{\psi}_{\sigma}(x)\bigr\rangle & = & \mathrm{Tr}[e^{T(L-\delta x)}(R_{\sigma}\otimes I)e^{\tilde{T}\delta x}(I\otimes\bar{R}_{\sigma'})]\nonumber\\
& & \qquad \times e^{i(q_{\sigma}-q_{\sigma'})x} e^{-iq_{\sigma'}\delta x},\nonumber\\
\bigl\langle\partial_{x}\hat{\psi}_{\sigma}^{\dagger}(x)\partial_{x}\hat{\psi}_{\sigma}(x)\bigr\rangle & = & \mathrm{Tr}\{e^{TL}[(iq_{\sigma}R_{\sigma}+\left[Q,R_{\sigma}\right])\otimes \mathrm{c.c.}]\},\nonumber\\
C_{\mathrm{pair}}(\delta x) & = & \mathrm{Tr}[e^{T(L-\delta x)}(R_{\uparrow}R_{\downarrow}\otimes I)e^{T\delta x}\nonumber\\
 &  & \qquad\times(I\otimes\bar{R}_{\uparrow}\bar{R}_{\downarrow})]e^{-i(q_{\uparrow}+q_{\downarrow})\delta x},\label{eq:expectation}
\end{eqnarray} where  $\tilde{T} = Q\otimes I+I\otimes\bar{Q}-R_{\uparrow}\otimes\bar{R}_{\uparrow}-R_{\downarrow}\otimes\bar{R}_{\downarrow}$, $C_{pair}(\delta x)\equiv\allowbreak \langle\psi_{\uparrow}^{\dagger}(x+\delta x)\psi_{\downarrow}^{\dagger}(x+\delta x)\psi_{\downarrow}(x)\psi_{\uparrow}(x)\rangle$, and  $\delta x\ge0$.   These expressions are independent of $x$ (with $\sigma'=\sigma$ in the first case) and, in general, translationally invariant expectation values are expected from our formalism whenever the operator conserves the total particle number and spin.  Likewise, similar expressions and conclusions can be drawn for $\delta x<0$.

The kinetic energy density, written in terms of $q_{\sigma}$, $Q$,
and $R_{\sigma}$, can be derived from a lattice-regularized expectation value $\langle\{\epsilon^{-1}\allowbreak[ \hat{\psi}_{\sigma}^{\dagger}(x_{n+1})-\hat{\psi}_{\sigma}^{\dagger}(x_{n})] \}\allowbreak\cdot\{\mathrm{h.c.}\}\rangle 
$ where $\epsilon\allowbreak\equiv x_{n+1}-x_{n}$.  It contains terms having powers of $\epsilon^{-1}$ and $\epsilon^{-2}$ that diverge in the continuum limit of $\epsilon\rightarrow 0$.  These divergent terms cancel out exactly in the bosonic cMPS, whereas the kinetic energy density expression needs to be regularized \cite{Haegeman2013a} in the fermionic cMPS. The regular expression for the kinetic energy density and the simplified form for $C_{\mathrm{pair}}(\delta x)$, both stated in Eqs. (\ref{eq:expectation}), follow after imposing the conditions:
\begin{equation}
\left\{R_{\uparrow},R_{\downarrow}\right\}=0\quad\mathrm{and}\quad R_{\sigma}^2=0.
\end{equation}
The optimal method to meet these conditions is to construct $R_{\uparrow}$ and $R_{\downarrow}$ to have special algebraic structures.  One systematic way to enforce the requirement $R_{\uparrow}^2=0$ is by having $R_{\uparrow}$ in a Jordan canonical form \cite{Horn2012} with only nontrivial Jordan blocks.  As the nilpotency degree of $R_{\uparrow}$ must be 2, all of its Jordan blocks must be $2\times2$, and we are lead to the unique choice $R_{\uparrow} = I_{D/2}\otimes \sigma_{+}/2$.  A general form for $R_{\downarrow}$ that satisfies both $R_{\downarrow}^2 = 0$ and $\left\{R_{\uparrow},R_{\downarrow}\right\}=0$ is
\begin{eqnarray}
R_{\downarrow} & = & [P_{D/2}^{-1}(I_{D/4}\otimes \sigma_{+}/2)P_{D/2}]\otimes\mathbf{\sigma}_{z}\\
 &  & +[P_{D/2}^{-1}(A_{D/4}\otimes \sigma_{+}/2+B_{D/4}\otimes I_{2})P_{D/2}]\otimes \sigma_{+}/2,\nonumber
\end{eqnarray}
where $A_{D/4}$, $B_{D/4}\in\mathbb{C}^{D/4\times D/4}$, and $P_{D/2}\in\mathbb{C}^{D/2\times D/2}$ is an invertible matrix.

Furthermore, we enforce $Q+Q^\dagger+R_{\uparrow}^\dagger R_{\uparrow}+R_{\downarrow}^\dagger R_{\downarrow}=0$  using the cMPS gauge freedom \cite{Verstraete2010,Haegeman2013a}, so that the general form of $Q$ becomes $Q =A-\frac{1}{2}R_{\uparrow}^{\dagger}R_{\uparrow}-\frac{1}{2}R_{\downarrow}^{\dagger}R_{\downarrow}$, where $A$ is an anti-Hermitian matrix that is independent of position.  Interestingly, in our construction of the $Q$ and $R_{\sigma}$ matrices, the $R_{\uparrow}$ matrix contains no variational parameters, and all parameters reside on the $Q$ and $R_{\downarrow}$ matrices.  All the information needed to correctly capture the physics of the spin-up particles is, nevertheless, correctly encoded in $Q$ and the transfer matrix $T$.\footnote{An additional useful parameter is $a>0$ used for scaling as $Q\rightarrow aQ$ and $R_{\sigma}\rightarrow \sqrt{a}R_{\sigma}$, which in the thermodynamic limit would leave $\mathrm{lim}_{L\rightarrow\infty}e^{TL}$ invariant \cite{Verstraete2010}.}

We note that our form of $R_{\sigma}$ naturally constrains the bond dimension $D$ of the cMPS to be a multiple of 4, and the total number of variational degrees of freedom between the $Q$ and $R_{\sigma}$ matrices becomes approximately $1.75D^2$.  For a given bond dimension $D$, it is desirable to retain the maximum number of nonredundant variational parameters, as the computational time required to compute the expectation values for an arbitrary state scales as $O(D^6)$.  Our implementation gives an improvement of nearly a factor of 2 as compared to a previous construction that had $D^2+D$ variational degrees of freedom \cite{Rispler2012}.  

\begin{figure}
\includegraphics[width=8.5cm]{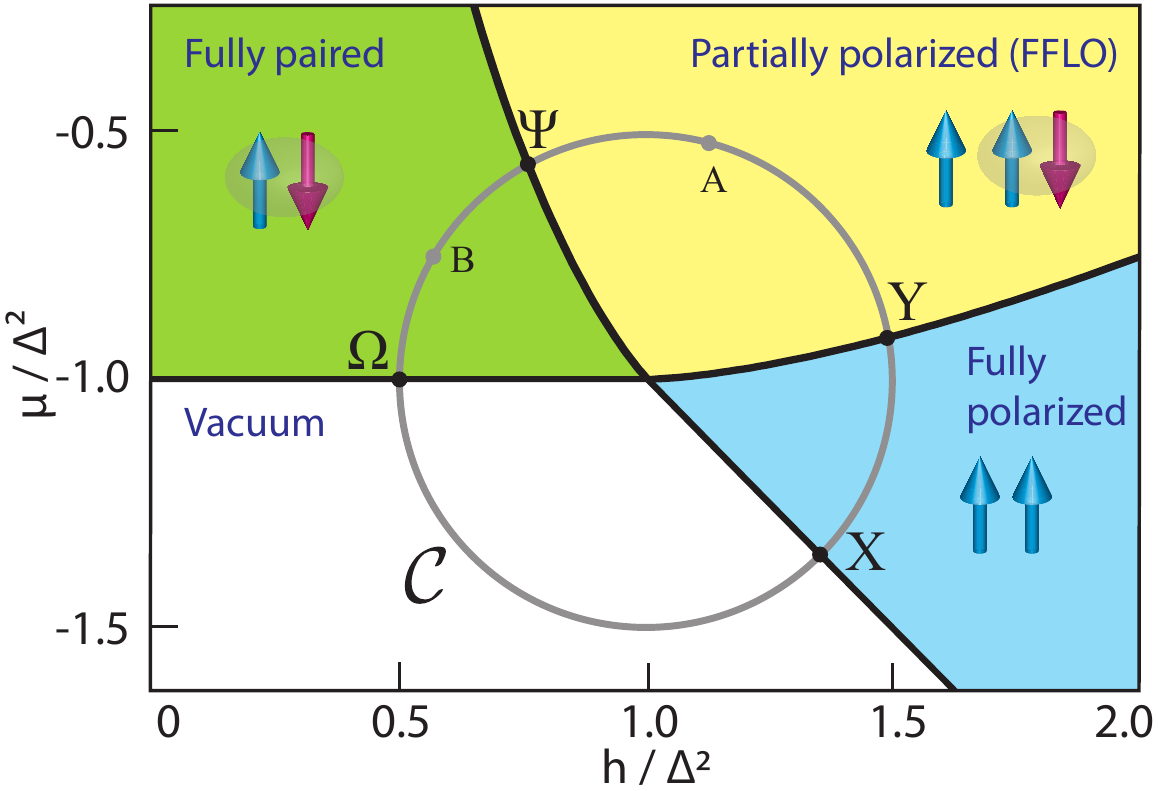}
\caption{\label{fig:phasediagram}
(Color online) Zero temperature ground-state phase diagram for the homogeneous attractive spin-1/2
Fermi gas in 1D. The phase boundaries were obtained from the Bethe ansatz \cite{Orso2007}.  The points X, Y, $\Psi$, and $\Omega$ along the circular trajectory $\mathcal{C}$ denote the intersections with the phase boundaries.  }
\end{figure}

In order to test our ansatz, we refer again to the Gaudin-Yang system in a grand canonical ensemble under the presence of a constant Zeeman field that is perpendicular to the system axis.  Figure \ref{fig:phasediagram} shows the ground-state phase diagram for the attractive Fermi gas in 1D, which has four distinct phases that are characterized by the population of each of the fermions: (i) fully polarized when
$n_{\uparrow}>n_{\downarrow}=0$ (between X and Y along the trajectory $\mathcal{C}$), (ii) partially polarized when $n_{\uparrow}>n_{\downarrow}>0$ (between Y and $\Psi$), 
(iii) fully paired when $n_{\uparrow}=n_{\downarrow}>0$ (between $\Psi$ and $\Omega$), and (iv) vacuum when
$n_{\uparrow}=n_{\downarrow}=0$ (between $\Omega$ and X).

To obtain variational ground states at the various phases, we optimized the expectation value of the
zero temperature free energy density $f \equiv F/L$ in the thermodynamic limit.  For the minimization, we implemented the simulated annealing algorithm \cite{Kirkpatrick1983}, taken as a variant of the simplex method that incorporates a Metropolis-like scheme \cite{Metropolis1953}.  The free energy density operator is $
\hat{f}=\hat{H}_{\mathrm{GY}}/L-\mu\left(\hat{n}_{\uparrow}+\hat{n}_{\downarrow}\right)-h\left(\hat{n}_{\uparrow}-\hat{n}_{\downarrow}\right)
$, where $\mu=(\mu_{\uparrow}+\mu_{\downarrow})/2$ is the
chemical potential, $h=(\mu_{\uparrow}-\mu_{\downarrow})/2$ the effective
magnetic field, and $n_{\sigma}$ the densities of the two species.
In terms of the variational matrices, we evaluated $\left\langle
f\right\rangle$ using the expressions in (\ref{eq:expectation}).

\begin{figure}
\includegraphics[width=8.7cm]{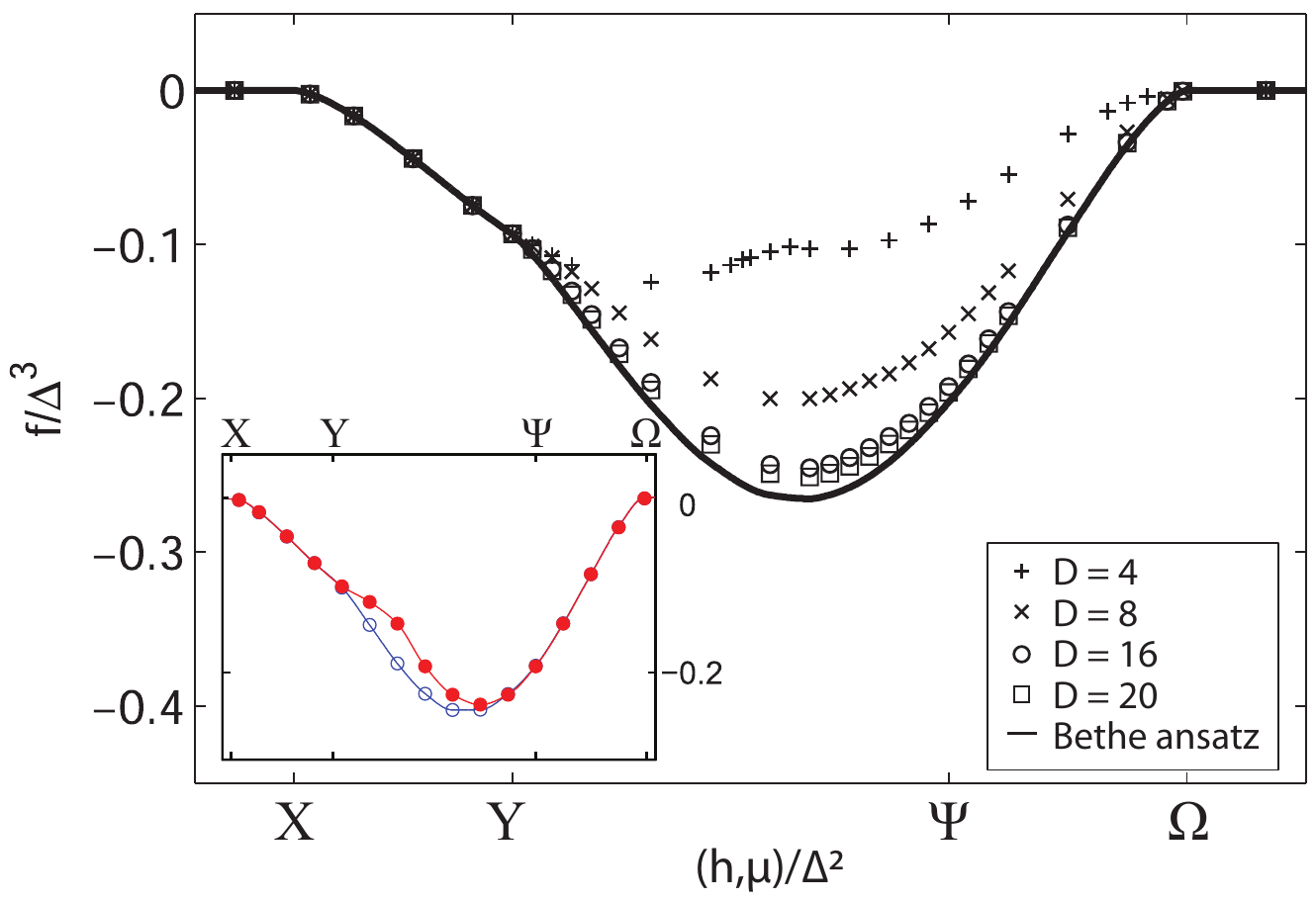}
\caption{\label{fig:freeenergy}
(Color online) Free energy densities along the trajectory $\mathcal{C}$
in Fig.~\ref{fig:phasediagram} obtained from variational states having
$D=4,8,16,20$ and the exact Bethe ansatz result (solid curve).  The inset has
free energies from variational states $(D=16)$ with
$\Rcal_{\sigma}(x)=R_{\sigma}$ (solid circles) and
$\Rcal_{\sigma}(x)=R_{\sigma}e^{iq_{\sigma}x}$ (open circles).}
\end{figure}

The free energy densities obtained from the optimization of the cMPS variational ansatz along the trajectory $\mathcal{C}$ are plotted in Fig.~\ref{fig:freeenergy}.  The bond dimensions range from $D=4$ to 20.  While a qualitative agreement starts to emerge from $D\geq8$, the variational ansatz with $D\geq16$ yields quantitatively accurate approximations, and convergence to the exact results is readily seen as $D$ is further increased.  The inset in Fig.~\ref{fig:freeenergy} presents the effect of neglecting the spatial modulation on the $\Rcal_{\sigma}(x)$ matrices by enforcing $q_{\uparrow}=q_{\downarrow}=0$.  In the partially polarized regime --- i.e. between the points Y and $\Psi$ --- we see notable gaps in free energy densities between the two curves, an indication that the modulation of $\Rcal_{\sigma}(x)$ is crucial for the accurate approximation of ground states in the partially polarized phase.  Moreover, we find that the relation $q_{\uparrow}/q_{\downarrow}=-n_{\downarrow}/n_{\uparrow}$ holds true in the partially polarized regime.

\begin{figure}
\includegraphics[width=8.5cm]{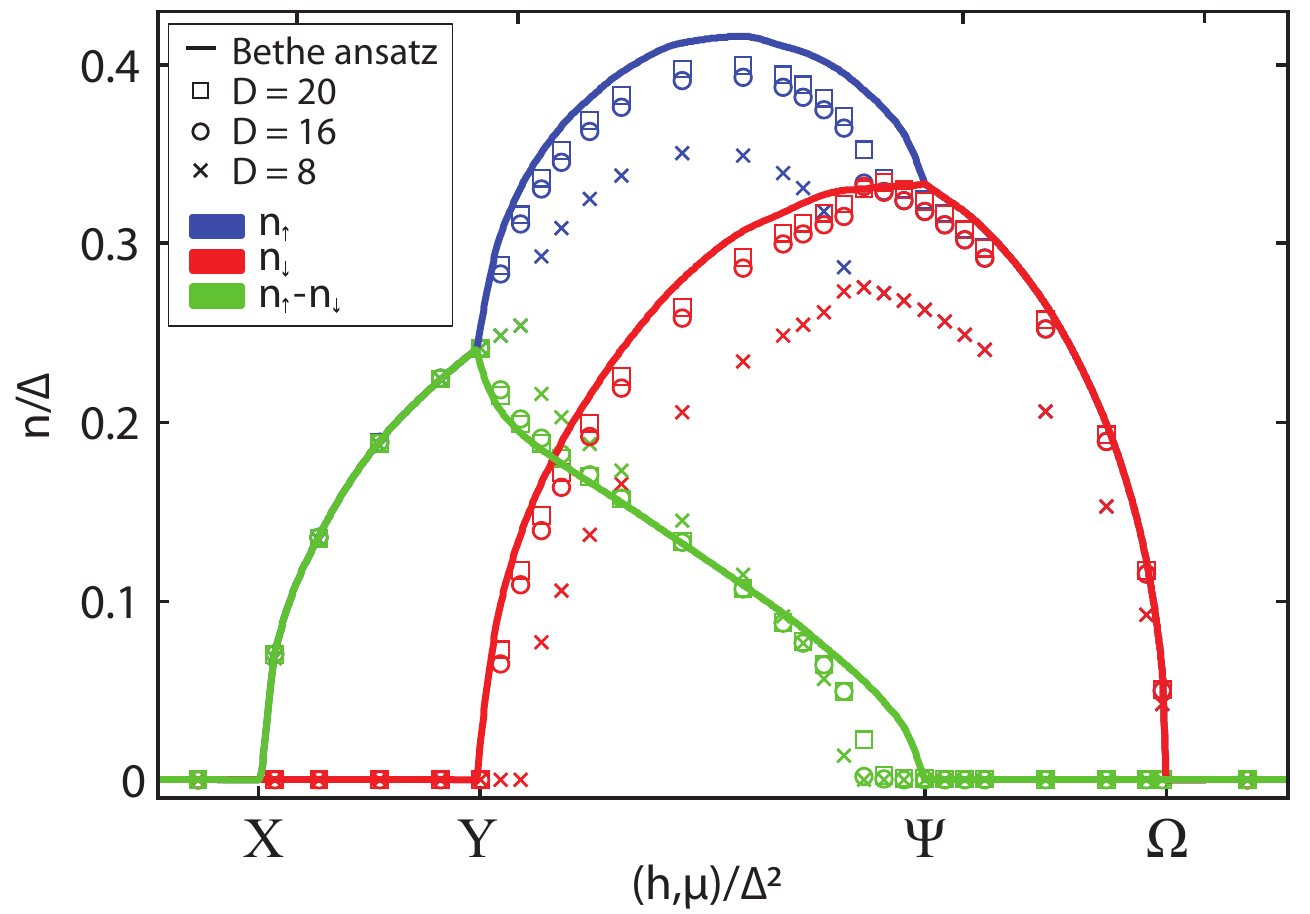}
\caption{\label{fig:density}
(Color online) Densities of the majority spins $n_{\uparrow}$ (blue),
minority spins $n_{\downarrow}$ (red), and their difference
$n_{\uparrow}-n_{\downarrow}$ (green) along the trajectory $\mathcal{C}$ in
Fig. \ref{fig:phasediagram}, as obtained from variational states having bond dimensions
$D=8,16,20$ and the exact Bethe ansatz result (solid curve).}
\end{figure}

Figure \ref{fig:density} is a plot of the densities $n_{\uparrow}$, $n_{\downarrow}$ and their difference along the trajectory $\mathcal{C}$.  As in the free energy density plot in Fig.~\ref{fig:freeenergy}, $D=8$ only gives results that qualitatively agree with the exact results, while $D\geq16$ show much more quantitatively reliable approximations.  The predictions for the phase boundary points X, Y and $\Omega$ are excellent, whereas the transition point $\Psi$ between the partially polarized and the fully paired phases converges more slowly; gradual improvement of the prediction with increasing $D$ is evident, nonetheless.

One of the exciting recent investigations by the ultracold-atom community is the effort to experimentally confirm whether the ground state of a spin-imbalanced spin-1/2 Fermi superfluid is, in fact, the Fulde-Ferrell-Larkin-Ovchinnikov (FFLO) state \cite{Fulde1964,*Larkin1965,*Casalbuoni2004}.  This state, postulated about 50 years ago, is characterized by a pair correlation function $\langle \hat{\psi}_{\uparrow}^{\dagger}(x+\delta x)\allowbreak\hat{\psi}_{\downarrow}^{\dagger}(x+\delta x)\allowbreak\hat{\psi}_{\downarrow}(x)\allowbreak\hat{\psi}_{\uparrow}(x)\rangle$ that has a long-range oscillatory behavior.  The renewed excitement was stirred by a theoretical prediction that a large portion of the ground-state phase diagram for an attractive 1D spin-1/2 Fermi gas is the FFLO phase (as seen in Fig.~\ref{fig:phasediagram}) \cite{Yang2001, Orso2007, Hu2007, Guan2007, Kakashvili2009}, paving an avenue for a feasible experimental verification \cite{Liao2010, Guan2013}.

\begin{figure}
\includegraphics[width=8.8cm]{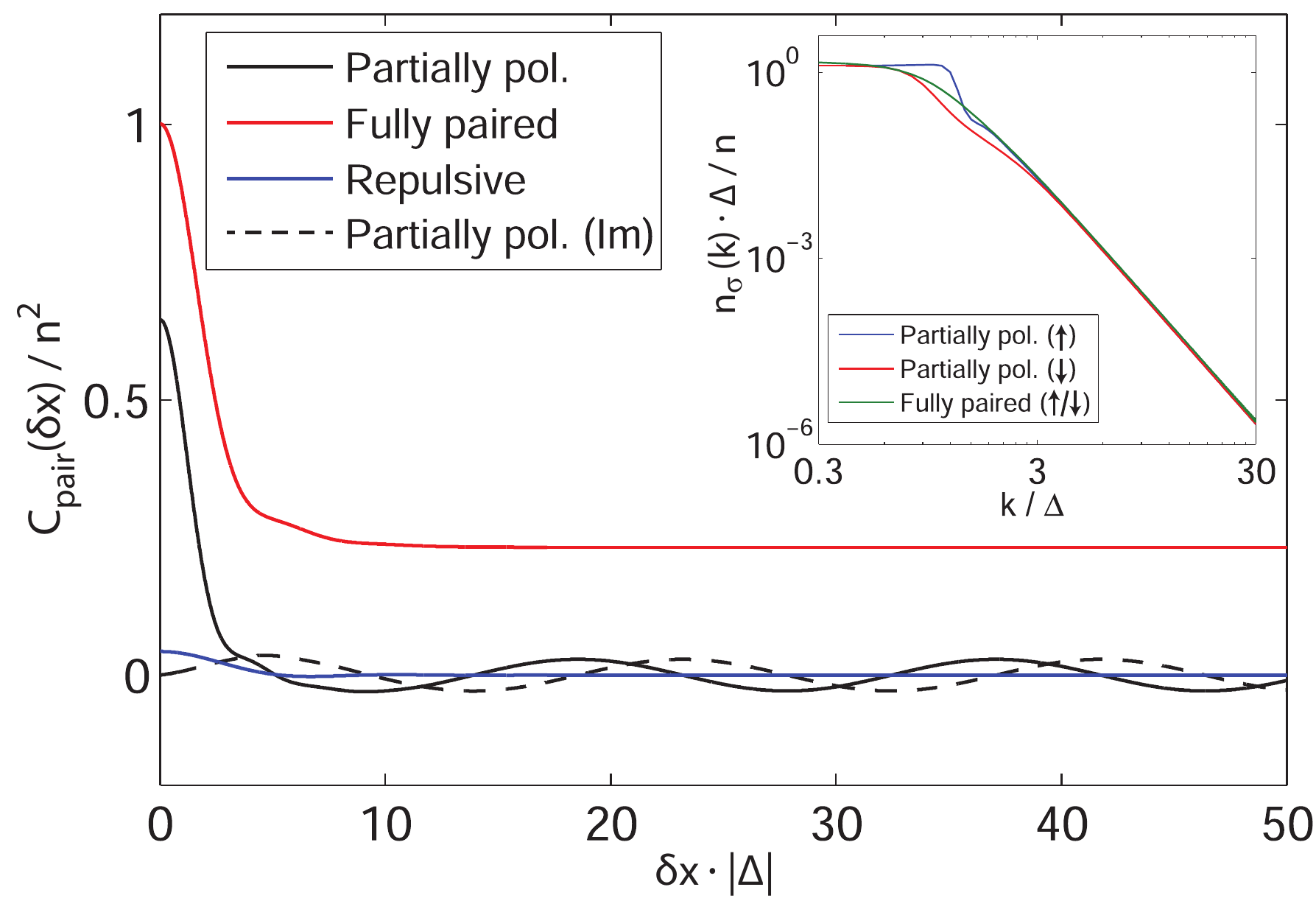}
\caption{\label{fig:pairpair}
(Color online) Real part of the pair correlation function in the
partially polarized phase (black) and the fully paired phase (red), evaluated
from variational cMPS $(D=16)$ at points A and B respectively (see Fig.~\ref{fig:phasediagram}). The black dashed curve is the imaginary part of the
pair correlation at A. The blue curve is obtained from a $D=16$ cMPS
in a repulsive system, with $\Delta=-1$, $\mu/|\Delta|^2=0.5$, and $h/|\Delta|^2=0.01$.
All correlators have been normalized by $n^2$, where
$n=n_{\uparrow}+n_{\downarrow}$. Inset: Fourier transform $n_{\sigma}(k)$ of $\langle
\hat{\psi}_{\sigma}^{\dagger}(x)\hat{\psi}_{\sigma}(0) \rangle$ evaluated at points A
and B, normalized by $n$. At large $k$, $n_{\sigma}(k)$ decays as $k^{-4}$ (cf.~Refs.~\cite{Haegeman2010,Casula2008}).}
\end{figure}

One of the advantages of an MPS formalism is its ability to express expectation values of essential correlations and observables in concise, closed forms.  Using the cMPS approximation of the ground state at point A in Fig.~\ref{fig:phasediagram}, we have evaluated the pair correlation function in the partially polarized regime via Eq.~(\ref{eq:expectation}).  In Fig.~\ref{fig:pairpair}, an oscillation that is present over a long distance is clearly visible, agreeing with the prediction that the ground state realizes the FFLO physics.  We note that this behavior was predicted beforehand, as $e^{T\delta x}$ converges to a fixed value for large $\delta x$ and the only remaining spatial dependence becomes $e^{-i(q_{\uparrow}+q_{\downarrow})\delta x}$.  On the other hand, the fully paired phase does not show an oscillatory pair correlation, and we verified there is no long-range correlation for the same system with a repulsive interaction. 

We have shown the validity of our implementation of a fermionic cMPS that incorporates a spatial modulation on the $\mathcal{R}_{\sigma}$ matrices with an efficient, explicit algebraic construction to meet the regularity conditions for the kinetic energy.  We have verified that our fermionic cMPS yields accurate results for interacting spin-1/2 Fermi gases in 1D (that converge to the exact results obtained from the thermodynamic Bethe ansatz), and have found that the variational states correctly predict essential observables and correlations, including a spontaneous translational symmetry breaking phenomenon, such as the FFLO superconductivity.  As the cMPS ansatz is generic and not restricted to integrable Hamiltonians, we envisage that our work would serve as a critical stepping stone towards solving outstanding 1D problems and shed light on our understanding of interacting Fermi fields.  Future study includes the extension of the cMPS to various interesting condensed-matter and cold-atom systems, such as spin-orbit coupled nanowires \cite{Lutchyn2010,*Oreg2010,*Stoudenmire2011,*Mourik2012,*Alicea2012,*Beenakker2013,*Roy2013,*Nadj-Perge2014}, FFLO superfluids in traps \cite{Casula2008,Liu2008,*Sun2011,Kakashvili2009} or tubular lattices \cite{Yang2001,Parish2007,*Zhao2008,*Feiguin2009,*Kim2012,*Sun2012,*Sun2013,*Sun2014}, and Bose-Fermi mixtures \cite{Viverit2000,*Albus2003,*Das2003}. 

\begin{acknowledgments}
We acknowledge discussions with J.~I.~Cirac and F.~Verstraete and the hospitality of the KITP at UCSB where those took place (NSF Grant No.~PHY05-51164). Funding for this work was provided by the University of Cincinnati and by the DARPA OLE program through ARO W911NF-07-1-0464; parallel computing resources were from the Ohio Supercomputer Center (OSC).
\end{acknowledgments}

\end{document}